\documentclass[letterpaper,11pt]{article}
\usepackage{graphicx,amssymb,amstext,amsmath,amsfonts}
\usepackage{fullpage}
\usepackage{cite}
\usepackage{bm}

\makeatletter
\def\@cite#1#2{\textsuperscript{{#1\if@tempswa , #2\fi}}}
\makeatother

\makeatletter
\def\@biblabel#1{#1.}
\makeatother

\newenvironment{methods}{%
    \section*{Methods}%
    \setlength{\parskip}{0pt}%
    }{}

\newcommand{\onlinecite}[1]{\hspace{-1 ex} \nocite{#1}\citenum{#1}}

\newcommand{\spacing}[1]{\renewcommand{\baselinestretch}{#1}\large\normalsize}
\spacing{2} 

\let\oldtitle=\title
\def\title#1{\oldtitle{\sffamily\bfseries{#1}}}
\let\oldauthor=\author
\def\author#1{\oldauthor{\sffamily\bfseries\normalsize{#1}}}

\def\@maketitle{%
  \newpage\spacing{1}\setlength{\parskip}{12pt}%
  {\Large\bfseries\noindent\sloppy \textsf{\@title} \par}%
    {\noindent\@author}%
}

\newenvironment{affiliations}{%
    \setcounter{enumi}{1}%
    \setlength{\parindent}{0in}%
    \slshape\sloppy%
    \begin{list}{\upshape$^{\arabic{enumi}}$}{%
        \usecounter{enumi}%
        \setlength{\leftmargin}{0in}%
        \setlength{\topsep}{0in}%
        \setlength{\labelsep}{0in}%
        \setlength{\labelwidth}{0in}%
        \setlength{\listparindent}{0in}%
        \setlength{\itemsep}{0ex}%
        \setlength{\parsep}{0in}%
        }
    }{\end{list}\par\vspace{12pt}}

\renewenvironment{abstract}{%
    \setlength{\parindent}{0in}%
    \setlength{\parskip}{0in}%
    \sffamily\bfseries%
    }{\par\vspace{-6pt}}

\title{Emergence of superconductivity from the dynamically heterogeneous insulating state in
La$_{\bm{2-x}}$Sr$_{\bm{x}}$CuO$_{\bm{4}}$}

\author{Xiaoyan Shi$^{1}$, G. Logvenov$^{2,3}$, A. T. Bollinger$^{2}$, I. Bo\v{z}ovi\'{c}$^{2}$, C. Panagopoulos$^{4,5}$ \& Dragana Popovi\'{c}$^{1*}$}

\begin{document}

\maketitle

\begin{affiliations}
 \item National High Magnetic Field Laboratory and Department of Physics, Florida State University, Tallahassee, Florida 32310, USA
 \item Brookhaven National Laboratory, Upton, New York 11973, USA
 \item Max Planck Institute for Solid State Research, Heisenbergstrasse 1, D-70569 Stuttgart, Germany
 \item Department of Physics, University of Crete and FORTH, GR-71003 Heraklion, Greece
 \item Division of Physics and Applied Physics, Nanyang Technological University, 637371 Singapore
\end{affiliations}
\begin{itemize}
  \item E-mail: dragana@magnet.fsu.edu
\end{itemize}

\newpage

\begin{abstract}
A central issue for copper oxides is the nature of the insulating ground state at low carrier densities and the emergence of high-temperature superconductivity from that state with doping.  Even though this superconductor-insulator transition (SIT) is a zero-temperature
transition, measurements are not usually carried out at low temperatures.  Here we use magnetoresistance to probe both the insulating state at very low temperatures and the presence of superconducting fluctuations in La$_{\bm{2-x}}$Sr$_{\bm x}$CuO$_{\bm 4}$ (LSCO) films, for doping levels that range from the insulator to the superconductor ($\bm{x=0.03-0.08}$).  We observe that the charge glass behavior, characteristic of the insulating state, is suppressed with doping, but it coexists with superconducting fluctuations that emerge already on the insulating side of the SIT.  The unexpected quenching of the superconducting fluctuations by the competing charge order at low temperatures provides a new perspective on the mechanism for the SIT.
\end{abstract}

In cuprates, the long-range-ordered antiferromagnetic (AF) ground state of the parent Mott insulator is destroyed quickly by adding charge carriers\cite{Kastner-review}.  The electronic ground state that separates it from a superconductor, which emerges at somewhat higher doping, remains poorly understood.  The high-temperature properties of this intermediate, ``pseudogap'' region have been studied extensively, in particular in the underdoped regime, i.e. on the superconducting side of the SIT.  For example, high magnetic fields that were applied to suppress superconductivity revealed the insulating nature of the underlying electronic state\cite{Ando-1995,GSB-1996}.  On the other hand, there are very few data at low temperatures, especially on the insulating side of the SIT.  In LSCO, it is known that, at low enough temperatures, the hole-poor, finite-size AF domains located in CuO$_2$ ($ab$) planes undergo cooperative freezing\cite{Cho92,Niedermayer98,magsusc-Lavrov} into an inhomogeneous, but magnetically ordered phase, often referred to as a cluster spin glass.  The doped holes seem to be clustered into areas that separate these AF domains\cite{Julien99,Singer02NQR,Dumm03EM,Ando02Ranisotropy,Ando03MR}.  In LSCO with $x=0.03$, they exhibit correlated, glassy behavior at even lower temperatures, deep within the spin-glass phase ($T \ll T_{SG}$)\cite{Ivana-PRL,Jelbert08,Ivana-pMR,Shi-PhysicaB}, suggestive of a charge glass transition as $T\rightarrow 0$.  The key question is how such an insulating, dynamically heterogeneous ground state evolves with doping and gives way to high-temperature superconductivity\cite{Vlad-Christos}.

On general grounds, the behavior near the zero-temperature SIT is expected to be influenced by quantum fluctuations.  In case of the electrostatically-induced SIT\cite{LSCO-SIT,YBCO-SIT}, the scaling analysis of the temperature dependence of the resistance $R(T)$ in LSCO near the critical doping $x_c\approx 0.06$ was interpreted\cite{LSCO-SIT} in terms of models\cite{MPAFisher} where Cooper pairs emerge already on the insulating side\cite{torque-Li,Diag-Li}: the transition is driven by quantum phase fluctuations and the localized pairs form a Bose glass.
However, one could question whether the extrapolation of the experimental results to low temperatures is accurate, and whether the effects of electrostatic doping are equivalent to those of chemical doping.  In this study, we use an independent and more direct technique to probe superconducting fluctuations and the properties of the insulating state near the SIT as a function of chemical doping; moreover, we extend the temperature range of measurements down to 0.3~K.

\begin{figure}
\centerline{\includegraphics[width=12cm]{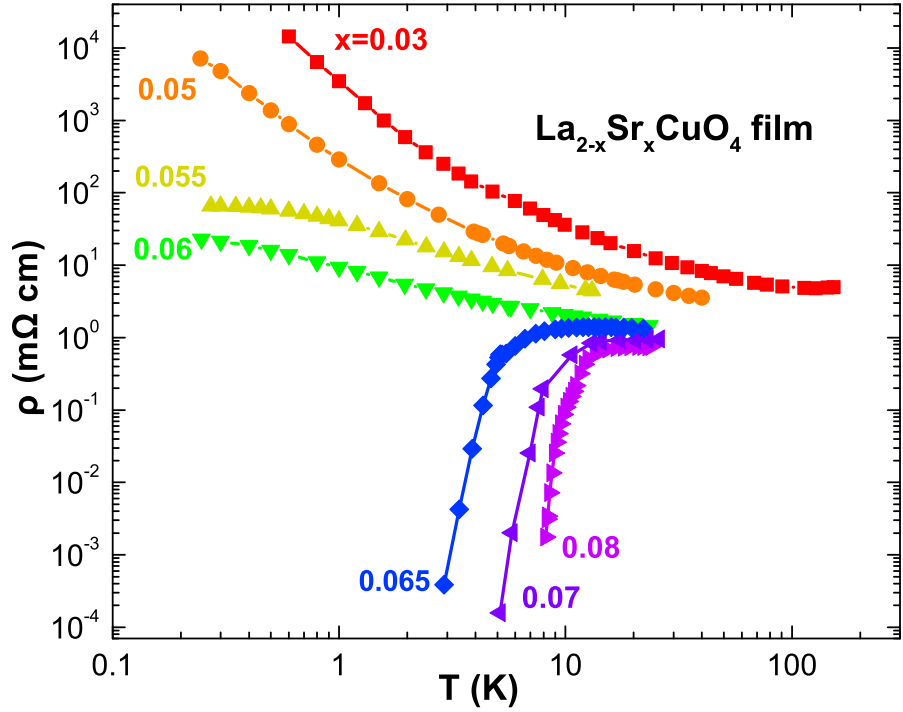}}
  \caption{\textbf{Temperature dependence of the in-plane resistivity $\rho$ for LSCO thin films with different doping levels $x$, as shown.} The critical temperatures $T_c$, defined as the transition midpoint (on a linear scale), are $(6\pm 1)$~K, $(9\pm 1)$~K and $(12\pm 1)$~K, for $x=0.065, 0.07, 0.08$ samples, respectively.}
  \label{fig:RvsT}
\end{figure}

The 100~nm thick films of LSCO were grown by atomic-layer-by-layer molecular beam epitaxy (ALL-MBE), which provides exquisite control of the thickness and chemical composition of the films\cite{MBE} (see Methods).  The samples with $0.03\leq x\leq0.06$ show insulating behavior in the in-plane $R(T)$, while those with $0.065\leq x\leq0.08$ become superconductors below the critical temperature $T_{c}(x)$ (Fig.~\ref{fig:RvsT}).  Compared to LSCO single crystals\cite{Ando-Tdep}, we find that the films are more resistive for the same nominal doping.
However, while insulating $R(T)$ for $x=0.03$ and 0.05 samples is described well by two-dimensional variable-range hopping\cite{Shi-PhysicaB}, the resistance increase with decreasing temperature is much weaker for $x=0.055$ and $x=0.06$ and cannot be fitted to any simple functional form.

\begin{figure}
\centerline{\includegraphics[width=14cm]{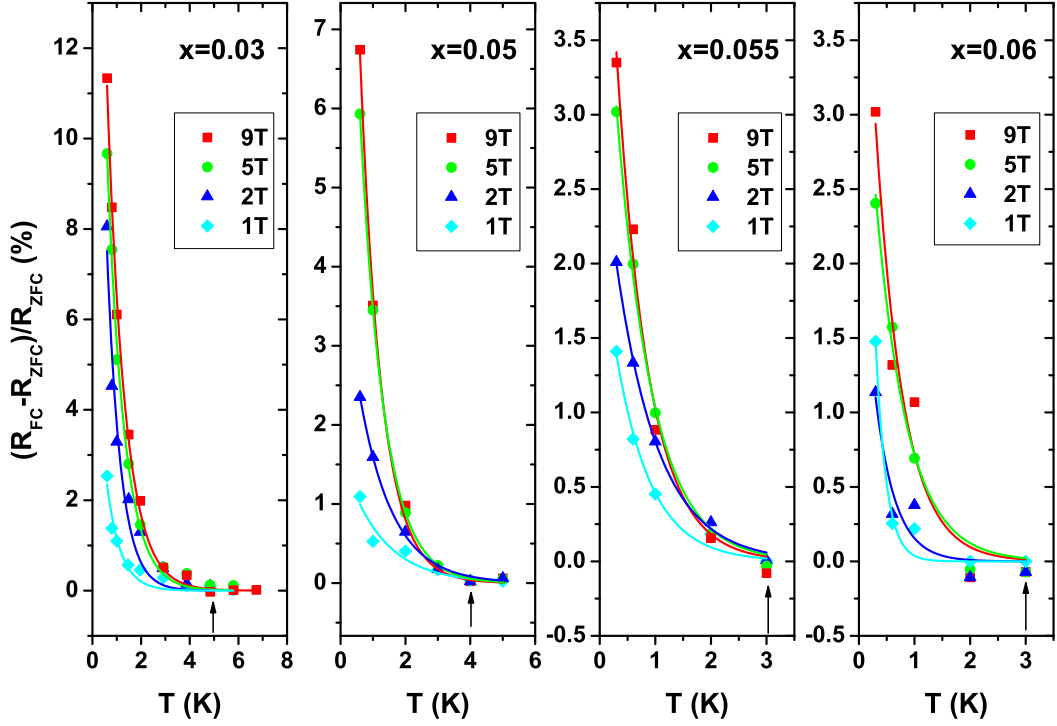}}
\caption{\textbf{The relative difference between field-cooled (FC) and zero-field-cooled (ZFC) ${\bm{R(H=0)}}$ in
LSCO films.}
The doping levels are $x=0.03, 0.05, 0.055$, and $0.06$, as shown.  In the FC protocol, the field was oriented perpendicular to CuO$_2$ planes and applied at $T>10$~K.  For each doping, $\mu_{0}H = 9$~T, 5~T, 2~T and 1~T were used during field cooling.  Arrows show $T^{\dagger}(x)$, the temperature where the difference between FC and ZFC values vanishes.  Solid lines are exponential fits to guide the eye.}
\label{fig:ZFC-FC}
\end{figure}

At the onset of the charge glass regime in LSCO single crystals with $x=0.03$, observed at $T \ll T_{SG}$, a difference appears between zero-field resistance $R(H=0)$ measured after zero-field cooling and cooling in a magnetic field\cite{Ivana-PRL,Ivana-pMR}.  This difference becomes more pronounced with decreasing temperature and it reflects the presence of frozen AF domains, such that only holes in the domain boundaries contribute to transport.  The magnetic field affects the domain structure because there is a weak ferromagnetic moment\cite{Thio}, oriented parallel to the $c$ axis, associated with each AF domain.  We find that all non-superconducting LSCO films exhibit such history dependence (Fig.~\ref{fig:ZFC-FC}) at $T<T^{\dagger}(x)$,
where $T^{\dagger}$ does not depend on the magnitude or the orientation of the magnetic field used during field cooling, but it decreases with doping.

\begin{figure}
\centerline{\includegraphics[width=15cm]{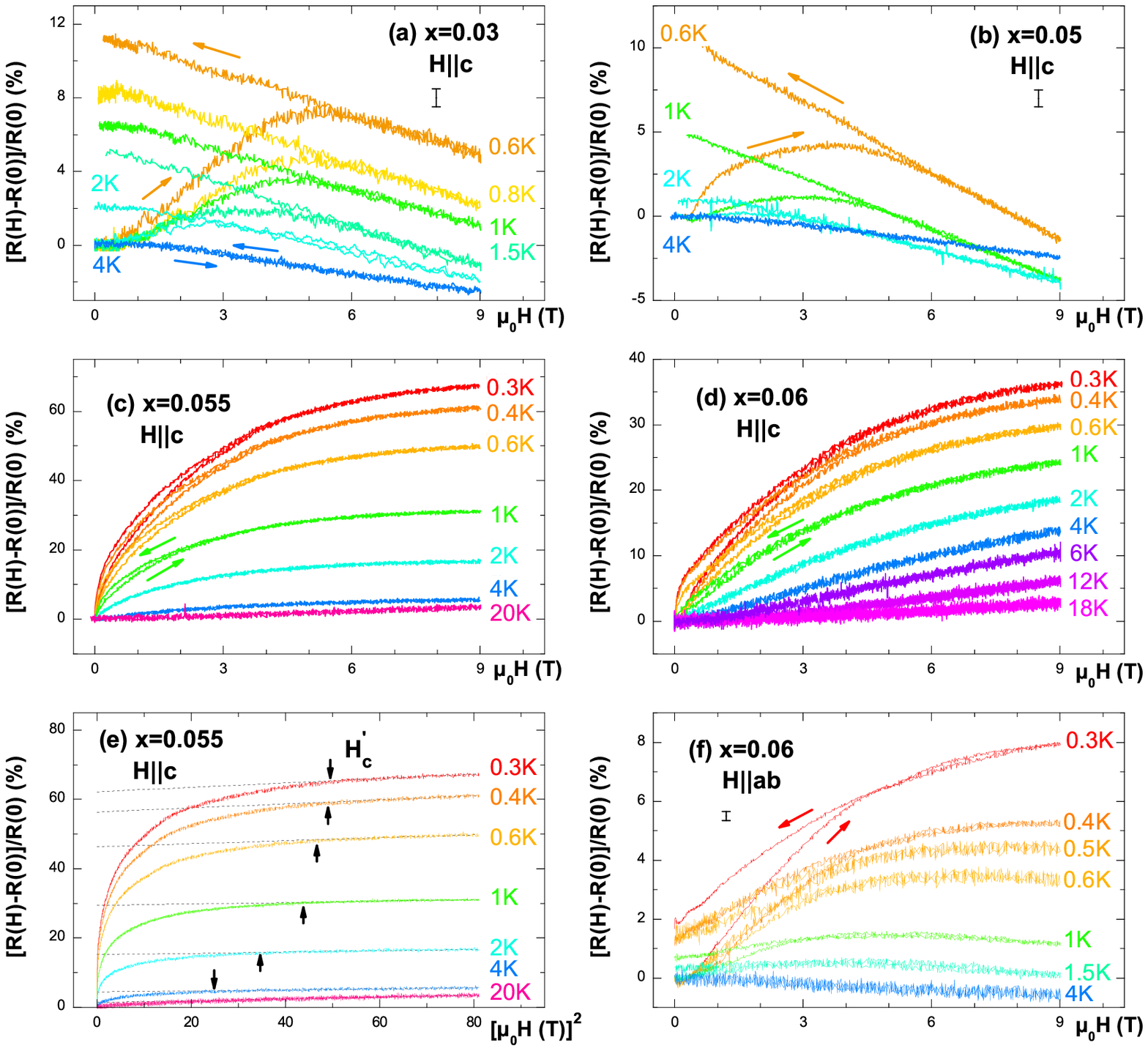}}
  \caption{\textbf{In-plane magnetoresistance of non-superconducting LSCO films at different temperatures.}  For each curve, the field was applied following zero-field cooling from $T\sim 10$~K.  The arrows show the direction of $H$ sweeps. Where shown, the error bars correspond to the typical change in the magnetoresistance due to temperature fluctuations.  The transverse ($H\parallel c$) magnetoresistance is shown for \textbf{a,} $x=0.03$, \textbf{b,} $x=0.05$, \textbf{c,} $x=0.055$ and \textbf{d,} $x=0.06$ doping levels.  \textbf{e,} The data from \textbf{c,} for sweep up, plotted \textit{vs.} $H^2$. Dashed lines are linear fits representing the contributions from normal state transport, \textit{i.e.} they correspond to $[R(H)-R(0)]/R(0) = [R_n(0)-R(0)]/R(0) +\alpha H^2$.  The intercept of the dashed line shows the relative difference between the fitted normal state resistance and the measured resistance at $H=0$.  Arrows show $H_c'$, the field above which superconducting fluctuations are fully suppressed and the normal state is restored.  \textbf{f,} The magnetoresistance for the $x=0.06$ film with field applied parallel to CuO$_2$ planes.  The sweep rate was 0.005~T/min for $\mu_{0}H<1$~T and 0.02~T/min for $\mu_{0}H>1$~T.}
  \label{fig:MR}
\end{figure}

Another manifestation of the onset of the charge glass behavior in strongly insulating, lightly doped La$_2$CuO$_4$ is the emergence of a hysteretic, positive magnetoresistance at low fields\cite{Ivana-PRL,Ivana-pMR,Shi-PhysicaB}.  In LSCO films that exhibit variable-range hopping transport, this effect is indeed observed at low enough temperatures (Fig.~\ref{fig:MR}a,b), giving rise to the history dependent
zero-field resistance and memory\cite{Shi-PhysicaB}.  The magnitude of the hysteretic, positive magnetoresistance is comparable for both $H\parallel c$ and $H\perp c$, as observed in single crystals\cite{Ivana-PRL,Ivana-pMR}.
A small increase in doping from $x=0.05$ to $0.055$, however, leads to dramatic changes in the magnetoresistance when the field is parallel to the $c$ axis (Fig.~\ref{fig:MR}c,d).  The magnetoresistance increases by almost an order of magnitude, and its positive component dominates in the entire experimental field range.  The hysteresis, however, is observed only over a limited range of the positive magnetoresistance, in contrast to the behavior in more insulating films (Fig.~\ref{fig:MR}a,b).  The results indicate that, in films with $x=0.055$ and $x=0.06$, another mechanism, most likely the suppression of superconducting fluctuations, also contributes to the positive magnetoresistance.  This is confirmed by measurements with field applied parallel to the $ab$ planes, which show that the non-hysteretic positive contribution is much weaker in that case (Fig.~\ref{fig:MR}f), as expected in the presence of superconducting fluctuations.

\begin{figure}
\centerline{\includegraphics[width=15cm]{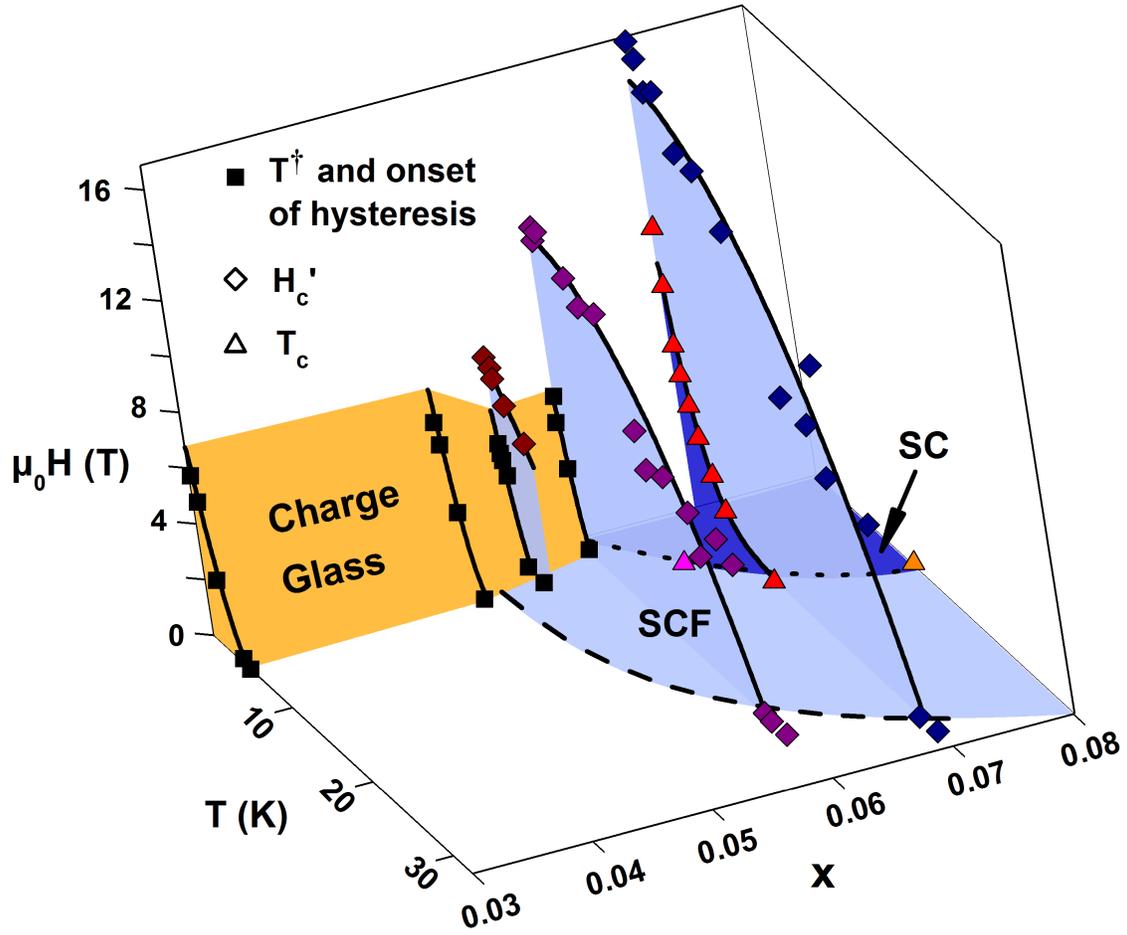}}
  \caption{\textbf{Phase diagram of the glassy region and of the onset of superconducting fluctuations in LSCO.}  Phase diagram shows the evolution of the glassy region and the emergence of superconducting fluctuations (SCFs) and superconductivity (SC) with doping, temperature and magnetic field.  The extent of the glassy regime does not depend on the field orientation. The range of SCFs is shown for the field applied perpendicular to CuO$_2$ planes.  Solid and dashed lines guide the eye.  Different colors of symbols for $H_{c}'(T)$ and $T_c(H)$ correspond to different values of doping. For both $x=0.06$ and $x=0.07$ films, $H_{c}'(T)=H_{c}'(0)[1-(T/T_2)^2]$ ($x=0.06$: $\mu_{0}H_{c}'(0)=11$~T, $T_2=24$~K ; $x=0.07$: $\mu_{0}H_{c}'(0)=15$~T, $T_2=29$~K.)}
  \label{fig:3D}
\end{figure}

Figure~\ref{fig:3D} shows the extent of the glassy region in temperature, field and doping, mapped out using the range of the hysteretic positive magnetoresistance, as well as the zero-field $T^{\dagger}(x)$ values.  Moreover, the extent of superconducting fluctuations can also be determined from the transverse ($H\parallel c$) magnetoresistance\cite{YBCO-SCF,YBCO-SCFlong,LSCO-SCF}.  In particular, above a sufficiently high magnetic field $H_{c}'(T)$, superconducting fluctuations are completely suppressed and the normal state is fully restored.  In the normal state at low fields, the magnetoresistance increases as $H^2$ (Ref.~\onlinecite{Harris}), so that the values of $H_{c}'$ can be found from the downward deviations from such quadratic dependence that arise from superconductivity when $H<H_{c}'$.  The magnetoresistance curves in Fig.~\ref{fig:MR}c,d indeed exhibit this kind of behavior, as illustrated in Fig.~\ref{fig:MR}e for the $x=0.055$ film.  A similar analysis of the data on the $x=0.06$ film at even higher magnetic fields is shown in Supplementary Figs.~1 and 2.  We note that the condition for the weak-field regime $\omega_c\tau\ll 1$ (with $\omega_c$ the cyclotron frequency and $\tau$ the scattering time) is easily satisfied for lightly doped LSCO films.  For $x=0.055$, for example, $\omega_c\tau\sim 0.01$ at $\mu_{0}H\sim 10$~T for $\rho\approx 1$~m$\Omega\cdot$cm  (see Fig.~\ref{fig:RvsT}).  The phase diagram constructed in Fig.~\ref{fig:3D} then shows, in addition to the glassy region, the values of $H_{c}'(T)$
and $T_c$ for different films, where $T_c$ was defined as the midpoint of the resistive transition.  The following conclusions may be drawn.

For strongly insulating $x=0.03$ and $x=0.05$ films, where superconducting fluctuations are not observed, the extent of the glassy behavior does not depend much on doping.  However, as glassiness is suppressed by further increase in doping, superconducting fluctuations (SCFs) emerge in insulating-like $x=0.055$ and $x=0.06$ films.  Here the fluctuations not only coexist with glassiness, but also they affect transport over a much wider range of $T$ and $H$ than glassy behavior.  At even higher doping, when superconductivity sets in, it is no longer possible to probe glassy dynamics using transport measurements.  While the role of superconducting fluctuations in the phenomenology of the pseudogap and their significance for understanding high-temperature superconductivity have been of great interest, there has been experimental disagreement about how high in temperature they may persist.  By tracking the restoration of the normal-state magnetoresistance, we find that the $H=0$ onset temperatures for SCFs in $x=0.055, 0.06$ and $0.07$ films (Fig.~\ref{fig:3D}) are lower by about $10-20$~K than those determined from the onset of diamagnetism\cite{Diag-Li} and Nernst effect\cite{Nernst} in LSCO crystals with similar $\rho(T)$ and $T_c$ values.
The origin of the discrepancy between onset temperatures for SCFs determined from different experiments, however, is still under debate\cite{Armitage,SCF-micro,SCF-Mott_Lara}.  There has been similar debate concerning the values of the upper critical field $H_{c2}(T\rightarrow 0)$ in LSCO and other cuprates. We note that $\mu_{0}H_{c}'(T=0)=(15\pm 1)$~T for the $x=0.07$ film is in agreement with $\mu_{0}H_{c2}\approx 16$~T obtained from specific heat measurements of a single crystal LSCO with a similar $T_c$ value\cite{SH-Wang}.  Specific heat results in LSCO at higher dopings\cite{SH-Wang} are, in turn, consistent with $H_{c2}(0)$ values determined from the $c$-axis resistive transport\cite{Ando-Hc2}.  Therefore, even though the method we employed to define $H_{c}'$
has an inherent limitation in accuracy, particularly at low temperatures where the $H^2$ dependence may be obscured by strong SCFs, we conclude that our determination of the onset of SCFs in a superconducting sample ($x=0.07$) with a low $T_c$, where the discrepancies between different techniques are less pronounced, is fairly consistent with other studies.  In non-superconducting samples with $x=0.055$ and $x=0.06$, the onset of SCFs takes place at even lower temperatures and fields (Fig.~\ref{fig:3D}), as expected.  We note that the $H_{c}'(T)$ line is well fitted by a simple quadratic formula $H_{c}'(T)=H_{c}'(0)[1-(T/T_2)^2]$ in both superconducting ($x=0.07$) and non-superconducting ($x=0.06$) samples.  The same $H_{c}'(T)$ dependence has been observed also in superconducting YBa$_2$Cu$_3$O$_y$ crystals\cite{YBCO-SCF,YBCO-SCFlong} and overdoped LSCO\cite{LSCO-SCF}.

\begin{figure}
\centerline{\includegraphics[width=15cm]{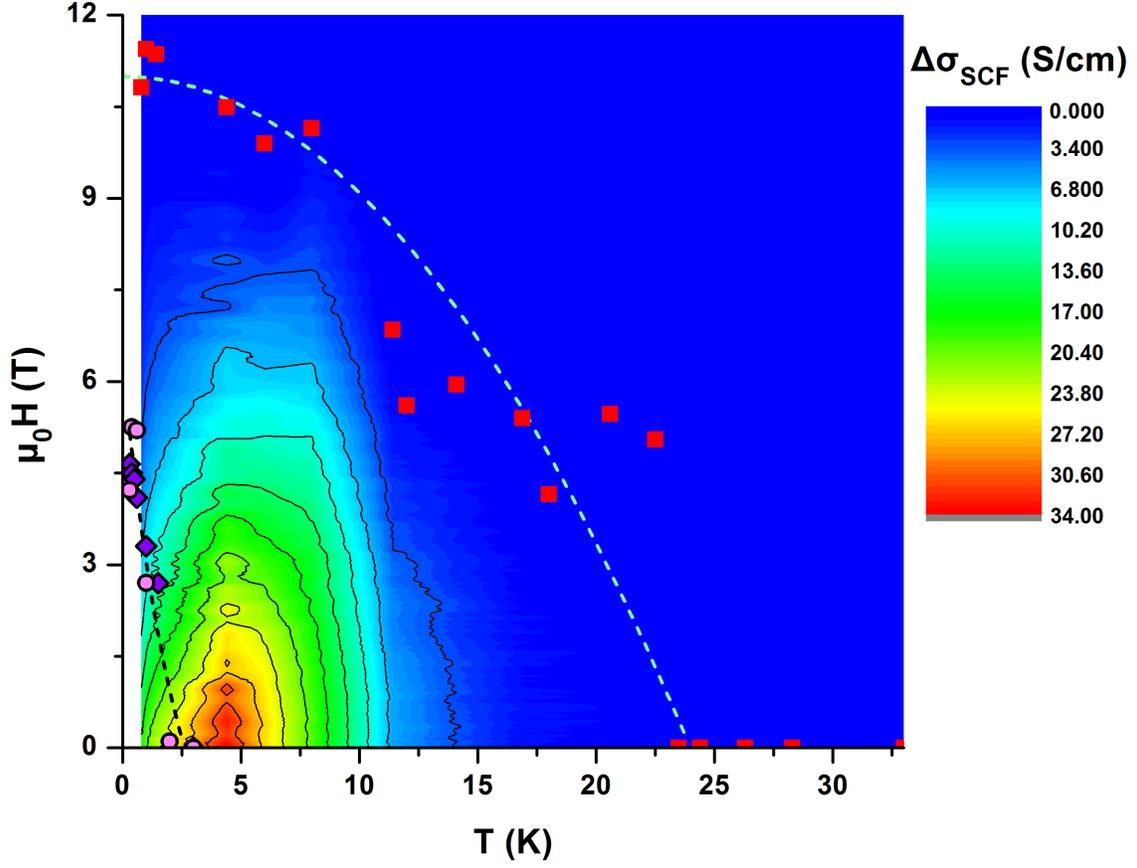}}
  \caption{\textbf{The contribution of superconducting fluctuations to conductivity and the glassy region in ${\bm{x=0.06}}$ LSCO film.}  The color map and contour plot shows the SCF contribution to conductivity $\Delta\sigma_{SCF}$ as a function of $T$ and $H\parallel c$.  Red squares represent $H_{c}'(T)$ and the green dashed line is a fit with $(\mu_{0}H_c')[$T$]=11[1-(T[$K$]/24)^2]$.  Pink dots ($H\parallel c$) and purple diamonds ($H\perp c$) show the extent of the charge glass region as determined from the measurements of the hysteretic positive magnetoresistance.}
  \label{fig:SCF}
\end{figure}

In order to explore the coexistence region in more detail, we calculate the SCF contribution to the conductivity\cite{YBCO-SCF,YBCO-SCFlong} $\Delta\sigma_{SCF}(T,H)=\rho^{-1}(T,H)-\rho_{n}^{-1}(T,H)$ using the measured resistivity $\rho(T,H)$ and the normal-state resistivity $\rho_{n}(T,H)$, where $\rho_{n}$ was obtained by extrapolating the region of $H^2$ magnetoresistance observed at high enough fields and temperatures (Fig.~\ref{fig:MR}e; see also Supplementary Figs.~1 and 2).  We emphasize that $\Delta\sigma_{SCF}(T,H)$ is not very sensitive to the values of $H_{c}'$, because the magnetoresistance at high fields is weak, i.e. the slope $\alpha$ of the $H^2$ dependence (see Fig.~\ref{fig:MR}e caption) is very small.  Therefore, all of our conclusions are qualitatively robust.  As shown in Fig.~\ref{fig:SCF} for the $x=0.06$ sample,
superconducting fluctuations are suppressed, as expected, by increasing temperature and magnetic field.  Quite unexpectedly, however, superconducting fluctuations are also suppressed at low temperatures, with the effect becoming stronger as temperature is reduced.
Similar behavior is observed in the $x=0.055$ film, but not in $x=0.07$, which becomes superconducting at $T_c=(9\pm 1)$~K.  This striking non-monotonicity in $\Delta\sigma_{SCF}(T)$ reveals the presence of a competing state.  By presenting the extent of the glassy region in $(T,H)$ on the same plot, it is clear that the competing state is precisely the dynamically heterogeneous charge order that is characteristic of the insulating phase.

Low-temperature experiments in very lightly doped La$_2$CuO$_4$ show that, as a result of long-range Coulomb interactions, holes form a collective, glassy state of charge clusters located in the CuO$_2$ planes\cite{Ivana-PRL,Jelbert08,Ivana-pMR,Shi-PhysicaB}.  Our results show that adding charge carriers in LSCO leads to the formation of localized Cooper pairs within this intrinsically heterogeneous charge ordered state, consistent with the Bose glass picture. By increasing the doping, the charge glass is suppressed, resulting in increased superconducting fluctuations, pair delocalization, and eventually the transition to a superconducting state.  Surprisingly, the superconducting fluctuations on the insulating side are quenched at low temperatures by the charge glass order.  Therefore, the pair localization and the onset of SIT in LSCO are influenced
by a competing charge order, and not merely by disorder, as it seems to be the case in some conventional superconductors\cite{Valles,Sacepe}.

The competition between charge order and superconductivity was revealed recently in YBa$_2$Cu$_3$O$_y$, a less disordered copper oxide, using nuclear magnetic resonance in the presence of high magnetic fields\cite{Julien-YBCO} that were required to destabilize superconductivity.  In contrast, our data show that the charge order in non-superconducting LSCO samples is present already in zero magnetic field.  These findings strengthen the idea that there is an intrinsic charge ordering instability in the CuO$_2$ planes.

\begin{methods}
The LSCO films were grown by atomic-layer-by-layer molecular beam epitaxy (ALL-MBE)\cite{MBE} on LaSrAlO$_4$ substrates with the $c$ axis perpendicular to the surface. The films were deposited at $T\approx 680~^{\circ}$C under $3\times 10^{-6}$~Torr ozone partial pressure. The growth was monitored in real-time by reflection high energy electron diffraction (RHEED) which showed that the films were atomically smooth and without any secondary-phase precipitates. The films are 75 unit cells (about 100~nm) thick; the measured $c = 1.312$~nm. Finally, a 160~nm thick gold layer was evaporated \textit{in situ} on top of the films for contacts.  The films were patterned using UV photolithography and ion milling to fabricate Hall bar patterns with the length $L = 2$~mm and the width $W = 0.3$~mm. The distance between the voltage contacts is 1.01~mm, and their width is 0.05~mm.  In order to remove any excess oxygen, the films were subsequently annealed in high vacuum ($4\times 10^{-5}$~Torr) for over an hour at $200-250~^{\circ}$C.

The in-plane sample resistance and magnetoresistance were measured with a standard four-probe ac method ($\sim11$~Hz) in the Ohmic regime, at $T$ down to 0.3~K realized in a $^3$He cryostat with magnetic fields up to 9~T and in the Millikelvin Facility at the National High Magnetic Field Laboratory with fields up to 18~T. The fields, applied either parallel or perpendicular to the CuO$_2$ planes, were swept at constant temperatures. The sweep rates, typically 0.02-0.03~T/min, were low enough to avoid the heating of the sample due to eddy currents. In both field orientations, the current $I\perp B$.
\end{methods}

\noindent\textbf{\large{Acknowledgements}}

\noindent We thank V. Dobrosavljevi\'c for discussions.  The work by X. S. and D. P. was supported by NSF/DMR-0905843 and the NHMFL, which is supported by NSF/DMR-0654118 and the State of Florida.  I.~B., G.~L. and A.~T.~B. were supported by the U.S. Department of Energy, Basic Energy Sciences, Materials Sciences and Engineering.  C.~P. was supported by EURYI, MEXT-CT-2006-039047, and the National Research Foundation, Singapore.

\noindent\textbf{\large{Author contributions}}

\noindent D.P. and C.P. conceived the project; the films were synthesized by ALL\_\,MBE and characterized by G.L. and I.B., and patterned by A.T.B.; X.S. designed the masks for lithography, helped with the patterning, performed the measurements, and analysed the data; X.S., I.B., C.P. and D.P. wrote the manuscript; D.P. planned and supervised the investigation.

\noindent\textbf{\large{Additional information}}

\noindent The authors declare that they have no competing financial interests.  Supplementary information accompanies this paper.  Correspondence and requests for materials should be addressed to D.P.~(email: dragana@magnet.fsu.edu).

\pagebreak

\begin{center}

\textbf{\Large{Supplementary Information for}}
\vspace*{6pt}

\textbf{\large{Emergence of superconductivity from the dynamically heterogeneous insulating state in
La$_{\bm{2-x}}$Sr$_{\bm{x}}$CuO$_{\bm{4}}$}}

\end{center}

\noindent Xiaoyan Shi$^{1}$, G. Logvenov$^{2,3}$, A. T. Bollinger$^{2}$, I. Bo\v{z}ovi\'{c}$^{2}$, C. Panagopoulos$^{4,5}$ \& Dragana Popovi\'{c}$^{1*}$

\begin{affiliations}
 \item National High Magnetic Field Laboratory and Department of Physics, Florida State University, Tallahassee, Florida 32310, USA
 \item Brookhaven National Laboratory, Upton, New York 11973, USA
 \item Max Planck Institute for Solid State Research, Heisenbergstrasse 1, D-70569 Stuttgart, Germany
 \item Department of Physics, University of Crete and FORTH, GR-71003 Heraklion, Greece
 \item Division of Physics and Applied Physics, Nanyang Technological University, 637371 Singapore
\end{affiliations}
\begin{itemize}
  \item E-mail: dragana@magnet.fsu.edu
\end{itemize}

\begin{figure}
\centerline{\includegraphics[width=14cm]{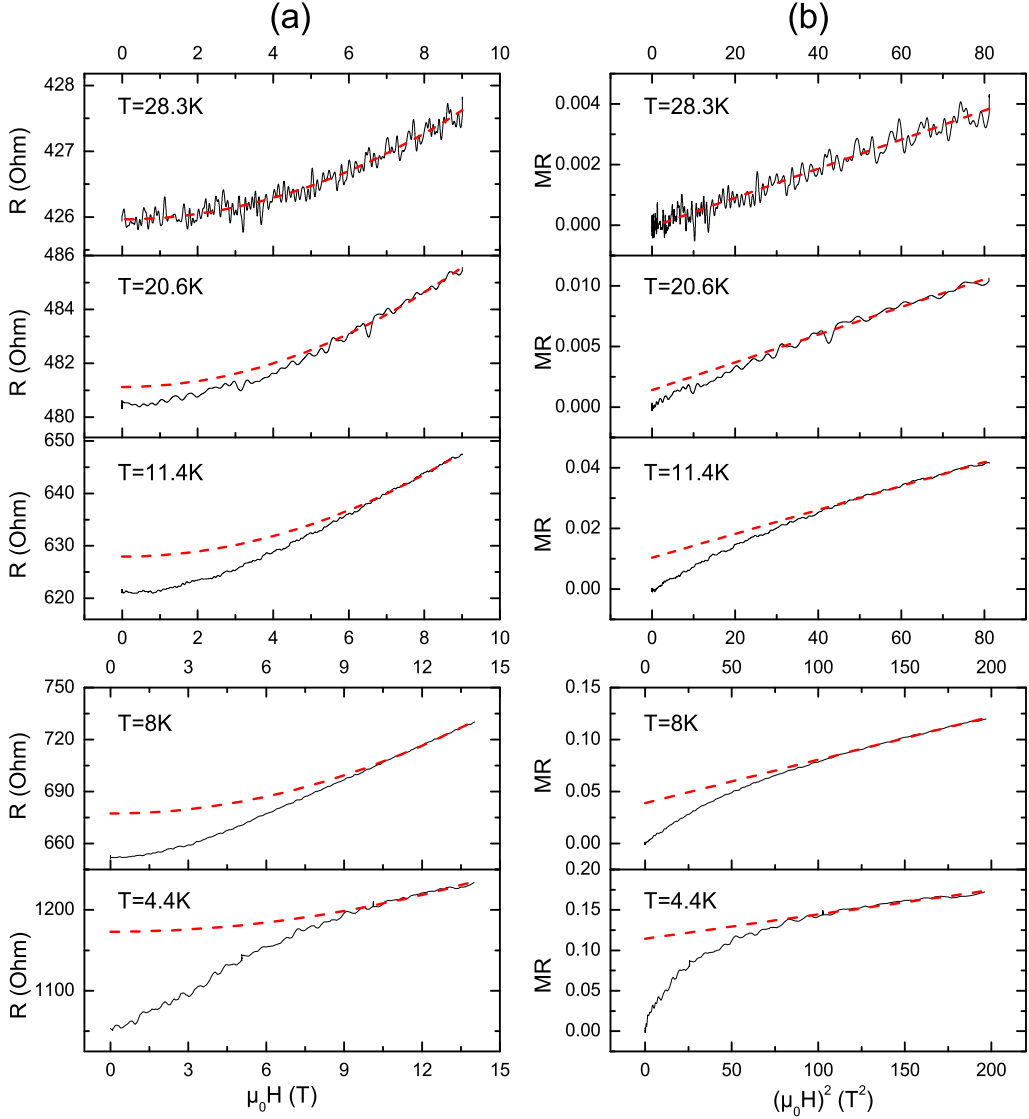}}
  \caption{\textbf{Transverse (${\bm{H\parallel c}}$) in-plane (a) resistance \textit{vs.} ${\bm{H}}$ and (b) magnetoresistance \textit{vs.} ${\bm{H^2}}$ for the ${\bm{x=0.06}}$ film at relatively high temperatures.} Dashed lines are fits representing the contributions from normal state transport. \textit{E.g.} in \textbf{b}, they correspond to $[R(H)-R(0)]/R(0) = [R_n(0)-R(0)]/R(0) +\alpha H^2$, where the intercept of the dashed line shows the relative difference between the fitted normal state resistance and the measured resistance at zero field.  The difference between the dashed lines and the measured magnetoresistance is due to the superconducting contribution.
  }
  \label{fig:MRfit}
\end{figure}
\begin{figure}
\centerline{\includegraphics[width=14cm]{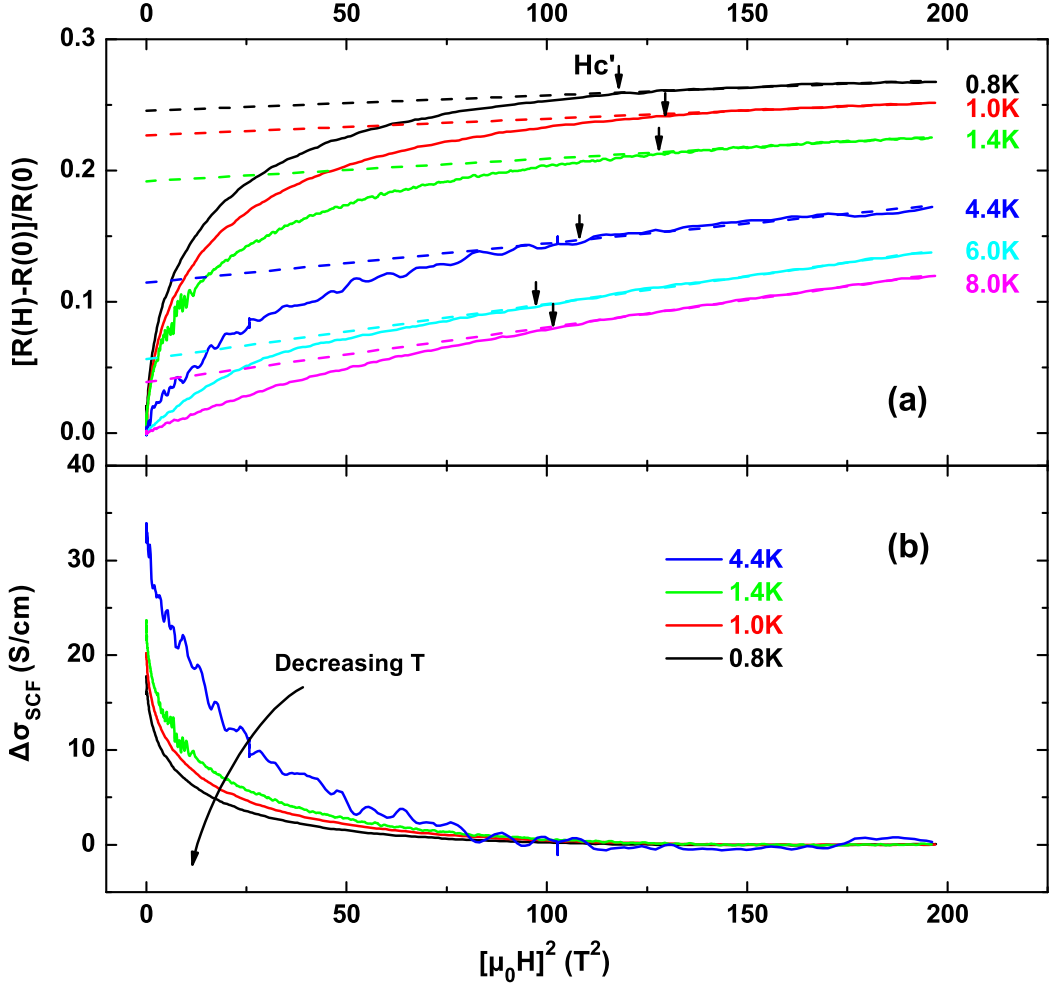}}
  \caption{\textbf{(a) Transverse (${\bm{H\parallel c}}$) in-plane magnetoresistance  and (b) the contribution of superconducting fluctuations to conductivity, ${\bm{\Delta\sigma_{SCF}}}$, \textit{vs.} ${\bm{H^2}}$ for the ${\bm{x=0.06}}$ film at low temperatures.} In \textbf{a}, dashed lines are fits representing the contributions from normal state transport, i.e. they correspond to $[R(H)-R(0)]/R(0) = [R_n(0)-R(0)]/R(0) +\alpha H^2$. The intercept of the dashed line shows the relative difference between the fitted normal state resistance and the measured resistance at zero field.  The difference between the dashed lines and the measured magnetoresistance is due to the superconducting contribution.  Arrows show $H_c'$, the field above which superconducting fluctuations are fully suppressed and the normal state is restored.  \textbf{b}, At the lowest temperatures ($T<5$~K), $\Delta\sigma_{SCF}$ decreases with decreasing temperature.}
  \label{fig:MRfit-low}
\end{figure}

\end{document}